\documentclass[fleqn,usenatbib]{mnras}

\usepackage{newtxtext,newtxmath}

\usepackage[T1]{fontenc}

\DeclareRobustCommand{\VAN}[3]{#2}
\let\VANthebibliography\thebibliography
\def\thebibliography{\DeclareRobustCommand{\VAN}[3]{##3}\VANthebibliography}

\usepackage{graphicx}	
\usepackage{amsmath}	
\usepackage{colortbl}
\usepackage{multirow}
\usepackage{subfigure}
\usepackage{xspace}
\usepackage{color}
\usepackage{url}
\usepackage{float}
\usepackage{ulem}
\usepackage{soul}
\usepackage{bm}
\usepackage{longtable}
\usepackage{threeparttable}
\usepackage{booktabs}
\usepackage{hyperref}
\usepackage{soul, xcolor}
\usepackage{lineno}

\title[The X-ray polarization of 4U 1822--37]{The First X-Ray Polarimetry of the Typical Accretion Disk Corona Source 4U 1822-37}

\author[Wu et al.]{
WanYun Wu,$^{1}$
Fei Xie,$^{1}$\thanks{E-mail: xief@gxu.edu.cn}
YuShan Ling$^{1}$
\\
$^{1}$Guangxi Key Laboratory for Relativistic Astrophysics, School of Physical Science and Technology, Guangxi University, Nanning 530004, China}

\date{Accepted XXX. Received YYY; in original form ZZZ}

\pubyear{\the\year{}}

\begin{document}
\label{firstpage}
\pagerange{\pageref{firstpage}--\pageref{lastpage}}
\maketitle

\begin{abstract}
We report the first Imaging X-ray Polarimetry Explorer (IXPE) observations of the typical accretion disk corona (ADC) source 4U 1822-37, covering 44 orbital cycles, together with simultaneous Swift-XRT data. The polarization degree (PD) of this source is found to be higher than typical low-mass X-ray binary (LMXB) values and exhibits a significant energy dependence. Within the 2--5~keV and 5--8~keV bands, PD remains constant, with average values of \(5.6\pm0.8\%\) and \(11.6\pm1.7\%\), respectively, and the difference is significant (\(>3\sigma\)). Orbital phase-resolved analysis reveals that during the companion eclipse, the PD in the 2--5~keV band drops from a significantly non-zero value to a level consistent with zero, while the polarization angle (PA) remains unchanged. In contrast, the PD in the 5--8~keV band shows a marginal rising trend, reaching \(21.3\pm6.1\%\), and the PA exhibits a periodic rotation of \(30^\circ\) before the eclipse. In addition, the PA rotates by \(40^\circ\) with increasing neutral absorption on a three-day timescale. These results suggest that the soft and hard emission in the 2--8~keV band may originate from regions with different spatial distributions, with the high PD implying a scattering contribution. The PA rotation is likely linked to scattering geometry asymmetries caused by a bulge or clumps.
\end{abstract}




\section{Introduction}  \label{sec:intro}

Low-mass X-ray binaries (LMXBs) are systems consisting of a compact object (a neutron star or a black hole) that accretes matter from its low-mass companion star via Roche-lobe overflow. For LMXBs containing a neutron star (NS-LMXBs), the X-ray radiation typically comprises a soft component and a hard component. The soft component originates from thermal radiation from the inner accretion disk or the neutron star surface, while the hard component arises from the comptonization of soft photons in a region of hot plasma. Additionally, relativistic jets and/or accretion disk winds are present in some LMXBs \citep{galaxies5040062,Allen2018TheDW}, further revealing the complex accretion geometry.

Periodic X-ray dips and/or eclipses have been observed in about 30 Galactic LMXBs \citep{Avakyan2023XRBcatsGL}, a phenomenon linked to the high inclination of these systems. The period of the dips is often consistent with the binary orbital period, and the X-ray intensity decreases by a smaller amount with a variable profile. This is interpreted as temporary obscuration of the central X-ray source by clumps of low-ionization material along the line of sight (see, e.g., \cite{Boirin2004AHA, Trigo2005SpectralCD, Shidatsu2013THEAD}). X-ray eclipses, on the other hand, occur when the X-ray emitting region is occulted by the companion star, causing the X-ray intensity to drop almost completely. However, residual flux can still be observed during the eclipse bottom in some eclipsing sources, providing evidence for the existence of a plasma region extending above the disk plane—namely, an extended accretion disk corona (ADC) \citep{White1981AccretionDC}, which is thought to form via dissipation processes within the inner disk or through evaporation of the disk surface material by the central X-ray source. In these sources, the outer disk with a certain vertical height obscures the central X-ray source, and the observed emission is mainly produced by electron scattering and reprocessing in photoionized media such as the corona.

\begin{table*}
\centering
    \begin{tabular}{ccccc}
        \toprule
        \midrule
              \textbf{Observatory}& \textbf{ObsID} & \textbf{Dates} & \textbf{Instrument} & \textbf{Exp. time (ks)}\\
        \midrule
        IXPE & 04003101 & 2025 Oct. 03 - 13& DU 1 &496.4\\
        & && DU 3 &496.5\\
        \midrule
        Swift-XRT & 00036691004 & 2025 Oct. 03& XRT-WT&1.8\\
        & 00036691005 & 2025 Oct. 05& XRT-WT&1.7\\
        & 00036691006 & 2025 Oct. 07& XRT-WT&1.7\\
        & 00036691007 & 2025 Oct. 09& XRT-WT&1.5\\
        & 00036691008 & 2025 Oct. 11& XRT-WT&1.6\\
        \bottomrule
    \end{tabular}
    \caption{Observations of 4U 1822-37 presented in the paper.}
\label{obs}
\end{table*}

X-ray polarimetry introduces a novel diagnostic dimension for studying NS-LMXBs. By measuring the polarization degree (PD) and polarization angle (PA), along with their variations with energy, time, and orbital phase, we can obtain constraints on the accretion geometry. The Imaging X-ray Polarimetry Explorer (IXPE) has observed several NS-LMXBs, revealing a rich variety of polarization behaviors (see the review by \citep{Ursini2024TheIV}). For instance, in Cir X-1, the polarization properties change significantly with the source state, including a PA rotation of up to 67\(^\circ\) and a marked evolution of PD between different branches, hinting at a possible misalignment between the neutron star's angular momentum and the orbital angular momentum \citep{Rankin2023XRayPV}. In the high-inclination dipping source GX 13+1 \citep{Bobrikova2024DiscoveryOA, DiMarco2025XRayDA}, the PA shows a swing of \(\sim\) 70\(^\circ\) between dipping and non-dipping states, which has been interpreted as a change in the fractional contribution of two spectral components, or the combined effect of scattering in the ADC or disk wind together with absorption by clumps, indicating that polarimetry is highly sensitive to the distribution of material along the line of sight.

4U 1822-37 is a particularly intriguing object among LMXBs. This source contains a neutron star, exhibiting clear X-ray pulsations with a period of \(P_{\rm s}=0.5924\) s \citep{Jonker2001DiscoveryOA}, and shows a significant (\(>99\%\) confidence) cyclotron resonant scattering feature (CRSF) at an energy of \(33\pm2\) keV with a depth of \(0.4^{+0.6}_{-0.3}\), which implies a potential magnetic field strength of \(4\times10^{12}\) G \citep{Sasano2013SuzakuVO}. These characteristics make it one of the few LMXBs that contains a strongly magnetized neutron star. 4U 1822-37 has a nearly circular orbit (eccentricity \(\sim0.03\)) with an orbital period of \(P\sim5.7\) hr \citep{White1981AccretionDC, Jonker2001DiscoveryOA}. The orbital inclination is constrained between 76\(^\circ\) and 84\(^\circ\) \citep{1989MNRAS.239..715H, Heinz2000TheFD}. Its orbital light curve displays a narrow and a broad intensity dip. The broad dip is interpreted as obscuration of the corona by a bulge at the outer edge of the accretion disk \citep{1989MNRAS.239..715H}, while the narrow dip is attributed to a partial eclipse of the corona by the companion star \citep{White1981AccretionDC, 1982ApJ...262..253M, 1989MNRAS.239..715H, 10.1093/mnras/258.3.457}. Furthermore, about 50\% of the X-ray flux remains detectable during the deepest phase of the eclipse, implying that the source is surrounded by an ADC with a size approximately 0.3 times the radius of the companion star \citep{White1981AccretionDC}. 

The distance of 4U 1822-37 is estimated to be \(\sim\) 2.5 kpc \citep{Mason1982PhaseRO, 2003AJ....125.2163C}, with an observed luminosity of \(\sim10^{36}\) erg s\(^{-1}\) \citep{1982ApJ...262..253M}. However, since the observed X-ray emission is believed to arise primarily from scattering or reprocessing in the photoionized medium, the intrinsic luminosity could be as high as \(\sim10^{37}\) erg s\(^{-1}\) \citep{Jonker2001DiscoveryOA, Sasano2013SuzakuVO}. The current understanding of the radiation geometry within the ADC of this system remains highly degenerate, with various phenomenological models capable of fitting the data (see, e.g., \cite{Heinz2000TheFD,Parmar2000BroadbandBO}). X-ray polarimetry can help probe the geometry and physical properties of different spatial regions within the source, thereby placing important constraints on the radiation models of ADC sources.

During the course of this work, \cite{Anitra2026TheFI} published an independent polarimetric study of 4U~1822--37 using the same IXPE observation, supplemented by Swift-XRT, XMM-Newton, and NuSTAR. Our results are consistent with theirs in terms of the average polarization parameters and their energy- and orbital-phase dependencies within the uncertainties. However, we adopted a smaller source region, which eliminates the need for additional background subtraction. In addition, we excluded possible faulty detector data and followed a standardized analysis pipeline using the \texttt{ixpeobssim} software and the PCUBE algorithm. This allowed us to reveal, for the first time, several new features, including a significant energy-dependent difference in PD, temporal variations of polarization parameters, and distinct orbital-phase dependencies of polarization parameters between the 2–5 keV and 5–8 keV bands. These findings have important implications for understanding the radiation structure of this source.

\begin{figure*}  %
\centering
\includegraphics[width=0.725\textwidth]{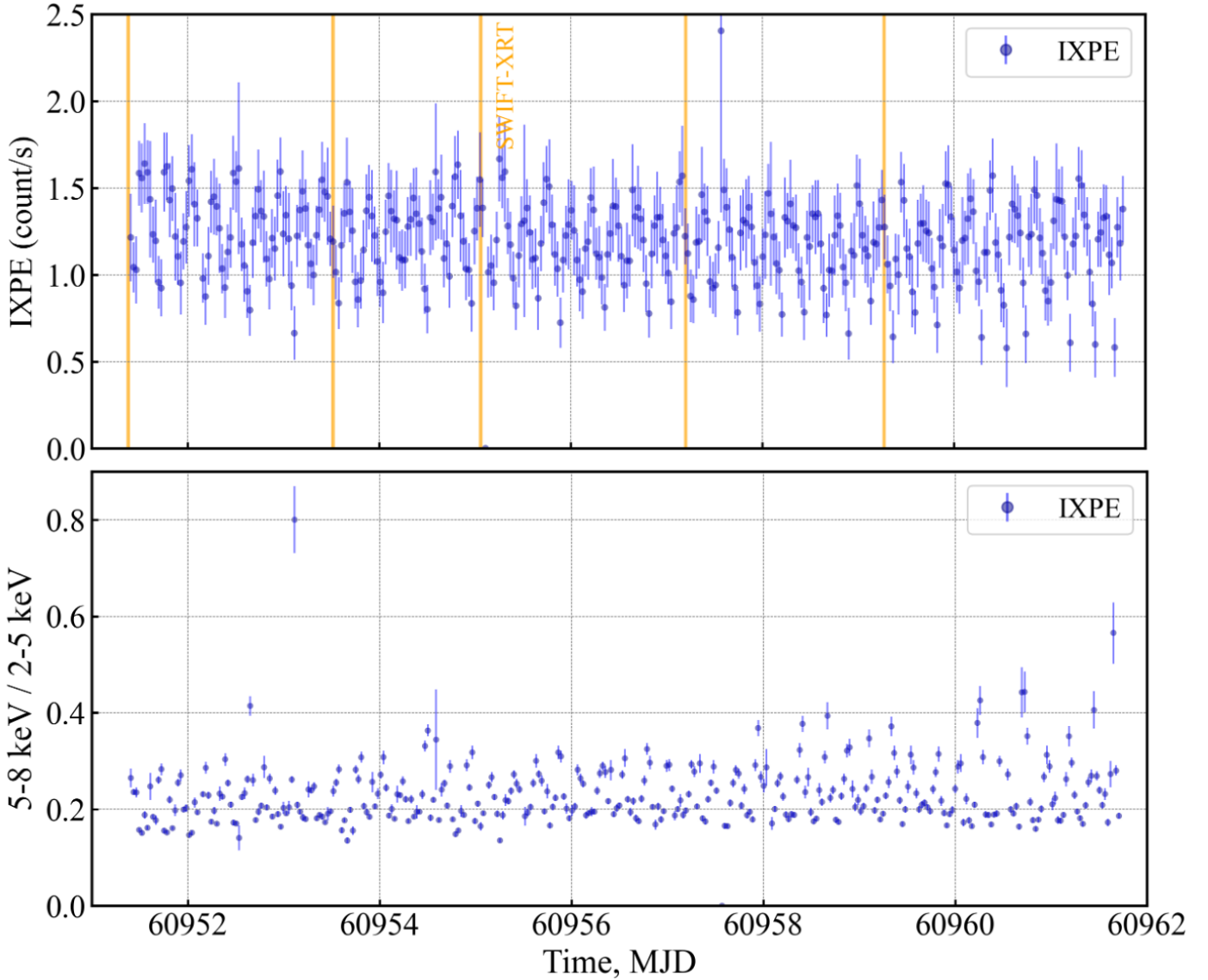}\\
\caption{Top panel: IXPE light curve of 4U 1822-37 in the 2--8~keV band, combining the two DUs, binned with a time interval of 2500~s. The orange shaded regions mark the Swift-XRT observation intervals. Bottom panel: Time evolution of the hardness ratio derived from the IXPE data, using the same binning as the light curve.}
\label{lc}
\end{figure*}

\begin{figure}  %
\centering
\vspace{5pt}
\includegraphics[width=0.485\textwidth]{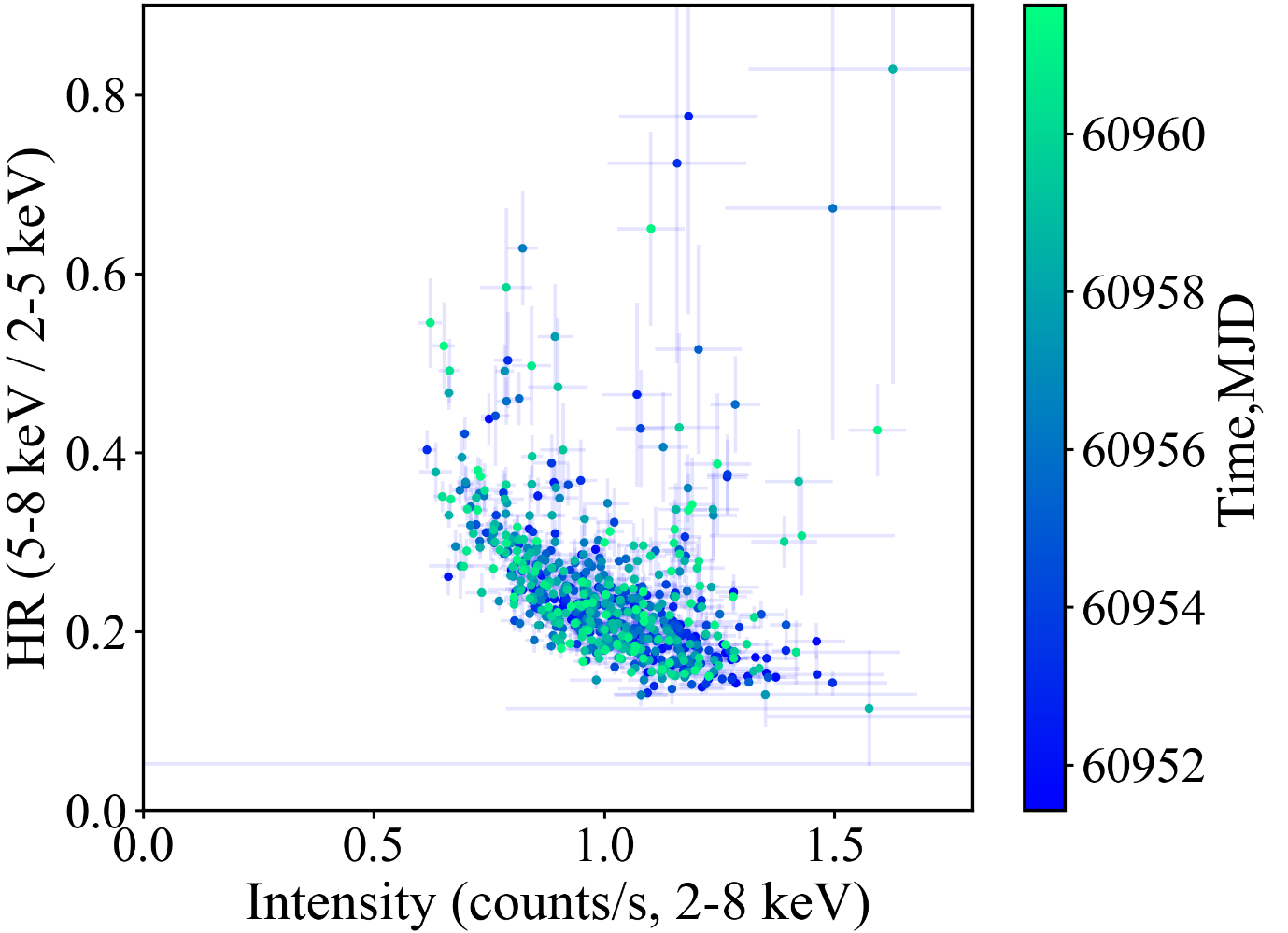}\\
\caption{Hardness-intensity diagram of 4U 1822-37 derived from the IXPE data. For plotting purposes, the data are rebinned to a bin width of 1000 s. The colors correspond to the evolution of the data with time, from blue at the beginning of the observation to green at the end of the observation.}
\label{time_hid}
\end{figure}

\section{Observation}  \label{sec:obs}

\subsection{IXPE}

IXPE is a mission jointly led by the National Aeronautics and Space Administration (NASA) and the Italian Space Agency, launched on 2021 December 9. Its detector system comprises three identical grazing-incidence telescopes, each of which comprises an X-ray mirror assembly and a polarization sensitive detector unit (DU), providing imaging, spectroscopy, and polarimetry measurements over the 2--8 keV energy band \citep{Soffitta2021TheIO, 2022AJ....164..103D, Weisskopf2022ImagingXP}.

IXPE observed 4U 1822-37 in a single pointing (ObsID 04003101) from 2025 October 3 to 13, covering approximately 44 orbital cycles, with a net exposure of \(\sim\) 496~ks per DU. During this observation, DU 2 was in a hardware anomalous state (after 2025 April 15, the ASIC reference voltage changed, some pixels in two rows failed, and the pixel noise increased). Although the IXPE team recalibrated DU 2 and released reprocessed data, given the potential residual systematic uncertainties of this detector, we used only stable DUs (DU 1 and DU 3) for the subsequent analysis\footnote[1]{See the IXPE team release: \url{https://heasarc.gsfc.nasa.gov/FTP/ixpe/data/obs/04/04003101/README.txt}}. Source events were extracted from the IXPE data using SAOImageDS9, selecting a circular region of radius 75\arcsec\ centered on the coordinates of 4U 1822-37. Given the high source flux, background subtraction was not performed, following the recommendation of \cite{DiMarco2023HandlingTB}.

Polarimetric analysis of the IXPE data was performed using the \textsc{ixpeobssim} software package (version 31.0.1; \cite{Baldini2022ixpeobssimAS}). We employed the PCUBE algorithm to generate cube polarimetry data and extracted the weighted energy spectra for the Stokes parameters $I$, $Q$, and $U$ using the PHA1, PHA1Q, and PHA1U algorithms, respectively \citep{DiMarco2022weights}. Subsequently, we binned the $I$, $Q$, and $U$ spectra to ensure a minimum of 30 counts per energy bin. Spectral fitting was performed within the \textsc{xspec} software package \citep{Arnaud1996}. All analyses utilized the IXPE calibration database (IXPE CALDB) released on 2024 February 28. In this paper, all uncertainties are quoted at the 68\% confidence level unless otherwise stated.

\begin{figure}
\centering
\vspace{5pt}
\includegraphics[width=0.485\textwidth]{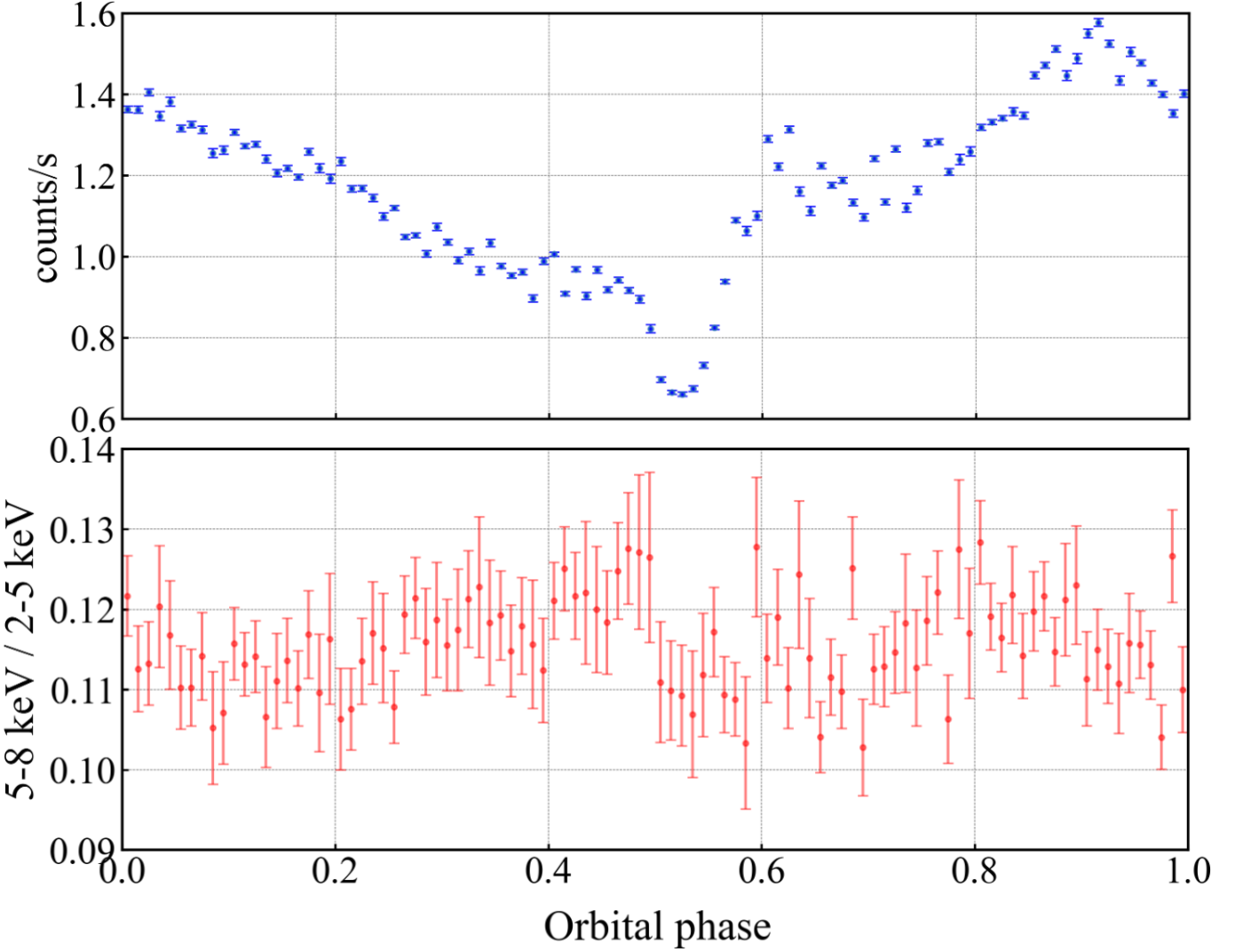}
\caption{Top panel: Orbital light curve of 4U 1822-37 folded at the 5.7~hr orbital period, derived from the IXPE data combining the two DUs. Bottom panel: The corresponding hardness ratio.}
\label{oblchr}
\end{figure}

\subsection{Swift-XRT}

During the IXPE observations, coordinated pointing observations of 4U 1822-37 were conducted with the X-Ray Telescope (XRT) aboard the Neil Gehrels Swift Observatory (Swift, \cite{2004ApJ...611.1005G}; see Table \ref{obs}). Swift-XRT is a focusing X-ray telescope operating in the energy range of 0.2--10 keV. This coordinated observation consisted of five separate snapshots, all performed in Windowed Timing (WT) mode.

Source and background events were extracted from the Swift-XRT data using the SAOImageDS9 software. Given that the brightness of 4U 1822-37 is far below the level that would cause significant central pixel pile-up effects \citep{Romano2006PanchromaticSO}, we employed a circular region with a radius of 80\arcsec\ centered on the coordinates of 4U 1822-37 to extract source events in order to collect more photons and improve statistical quality, rather than using an annular region. 
\begin{table*}
\centering
    \begin{tabular}{c|ccccc}
        \toprule
        \hline
             \textbf{E} & \textbf{PD} & \textbf{PA} & \(\bm{Q/I}\) & \(\bm{U/I}\) & \(\bm{\mathrm{MDP_{99}}}\)\\
             \textbf{(keV)} & \textbf{(\%)} & \textbf{(deg)} & \textbf{(\%)} & \textbf{(\%)} & \textbf{(\%)}\\
        \midrule
        2-3 & 5.1\(\pm\)1.3 & -17\(\pm\)7 & 4.2\(\pm\)1.3 & -2.8\(\pm\)1.3 & 4.0\\
        3-4 & 6.4\(\pm\)1.2 & -17\(\pm\)5 & 5.3\(\pm\)1.2 & -3.5\(\pm\)1.2 & 3.5\\
        4-5 & 5.5\(\pm\)1.5 & -26\(\pm\)8 & 3.5\(\pm\)1.5 & -4.3\(\pm\)1.5 & 4.5\\
        5-6 & 12.7\(\pm\)1.9 & -19\(\pm\)4 & 9.9\(\pm\)1.9 & -7.9\(\pm\)1.9 & 5.8\\
        6-7 & 14.2\(\pm\)2.6 & -26\(\pm\)5 & 8.7\(\pm\)2.6 & -11.3\(\pm\)2.6 & 7.8\\
        7-8 & 7.4\(\pm\)4.5 & -14\(\pm\)17 & 6.5\(\pm\)4.5 & -3.5\(\pm\)4.5 & 13.6\\
        \midrule
        2-8 & 7.8\(\pm\)0.9 & -20\(\pm\)3 & 5.9\(\pm\)0.9 & -5.1\(\pm\)0.9 & 2.8\\
        \bottomrule
    \end{tabular}
    \caption{Polarization properties in different energy bands obtained by PCUBE algorithm. \(\mathrm{MDP_{99}}\) is the minimum detectable PD at the 99\% confidence level.}
\label{energypd}
\end{table*}

\begin{figure*}  %
\centering
\hspace{2pt}
\includegraphics[width=0.46\textwidth]{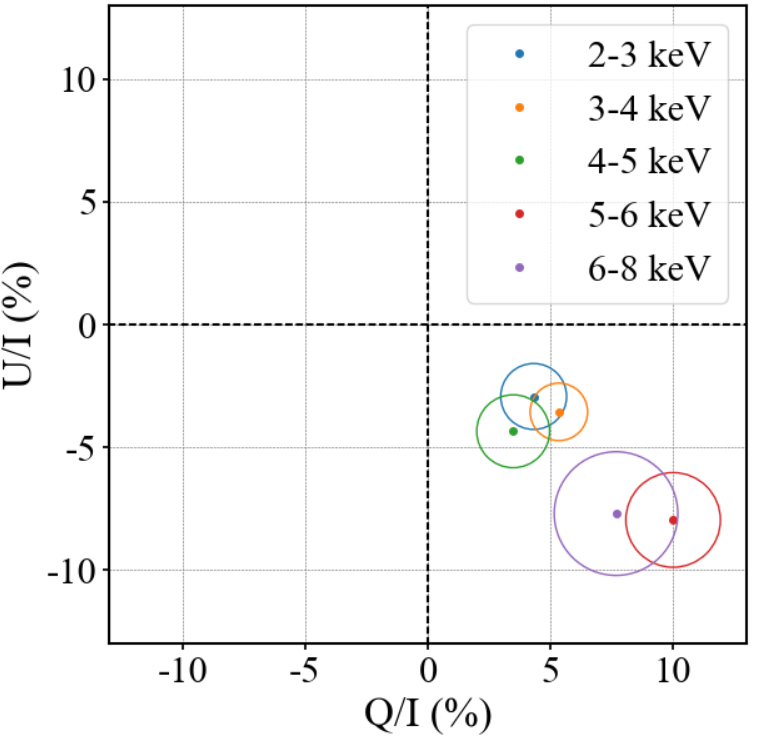}%
\hspace{15pt}
\includegraphics[width=0.459\textwidth]{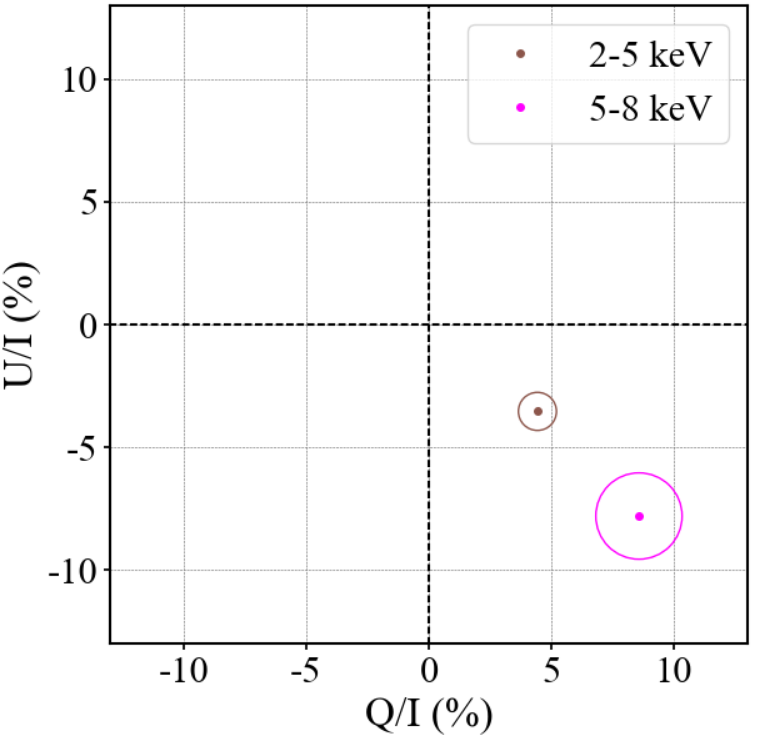}\\
\caption{Energy dependence of the normalized Stokes parameters Q/I and U/I obtained by the PCUBE algorithm.}
\label{energyqi}
\end{figure*}
Background events were extracted from a circular region offset by approximately 80\arcsec\ from the source. The Swift-XRT spectral data were fitted in the 0.5--8 keV band to optimize statistics, and the spectra were binned to ensure a minimum of 20 counts per energy channel. The analysis utilized the version of the Swift calibration database (Swift CALDB) released on 2023 July 25.

\section{Results}  \label{sec:Results}

We plot the IXPE light curve and the hardness ratio (HR) curve (5--8 keV/2--5 keV) in Figure \ref{lc}. The hardness-intensity diagram (HID, Figure \ref{time_hid}) shows that during the IXPE observation, the source shows little state variability. The counts and HR tracks over time are distributed almost uniformly within a region, indicating little variability between individual orbits.

We folded the orbital light curve using the orbital period \(P_{\rm orb} = 0.23210887(15)\) d and its derivative \(\dot{P}_{\rm orb} = 1.3(3) \times 10^{-10}\) d d\(^{-1}\) (at \(T_0 =\) MJD 45614), which are taken from \cite{Jain2010NewMO} based on observations covering 13 years with RXTE-PCA, RXTE-ASM, Swift-XRT, XMM-Newton, and Chandra. Given the large time span of the extrapolation from \(T_0\) to the IXPE observation date, we estimated the effect of orbital parameter uncertainties on the folded orbital phase, which does not produce a discernible impact on the profile of the orbital light curve during the IXPE observation. Furthermore, to better visualize the eclipse structure in the plot, we shifted the phase zero to the right by 0.28 after folding the light curve, and adopted this shifted value throughout our analysis. The resulting IXPE 2--8 keV orbital light curve and orbital HR curve are shown in Figure~\ref{oblchr}, and are consistent with previous results in the soft energy band \citep{Parmar2000BroadbandBO, Jain2010NewMO}. The source count rate declines slowly in orbital phase 0.90--1.00 and 0.00--0.45, then drops steeply by 50\% at phase 0.45--0.60, a phase interval that is considered to be the companion eclipse, while the HR drops sharply at phase 0.50--0.60 by \(\sim\) 28\%. After the eclipse, the count rate rises slowly, while the HR fluctuates within a certain range.

\begin{figure}  %
\centering
\vspace{3pt}
\includegraphics[width=0.485\textwidth]{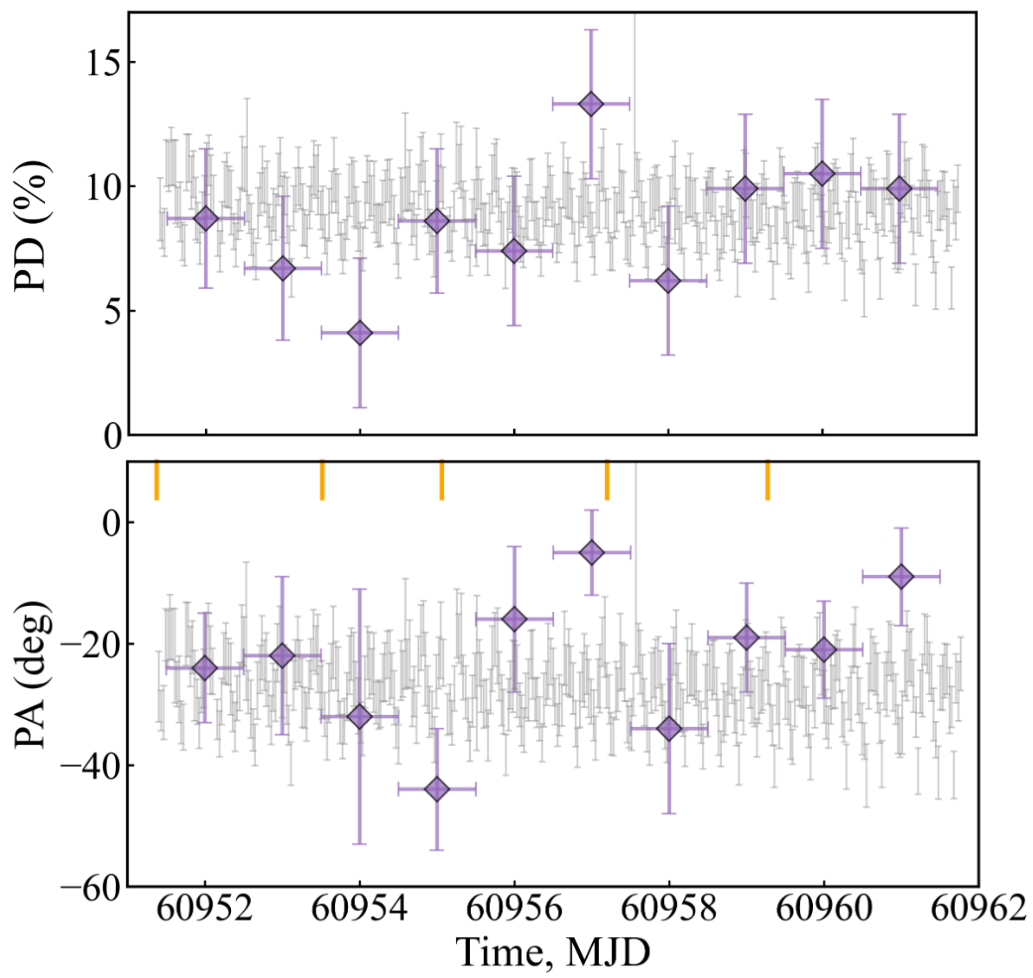}\\
\caption{Top panel: PD as a function of time during the IXPE observation obtained by the PCUBE algorithm. Bottom panel: PA as a function of time. The gray error bars show the profile of the IXPE light curve. The orange regions in the bottom panel mark the Swift-XRT observation intervals.}
\label{time_pdpa}
\end{figure}

\subsection{Polarimetric analysis} \label{Polarimetric analysis}

Using IXPE data in the 2--8\,keV energy band for polarization analysis, we obtained an average PD of \( 7.9\pm0.9\%\) and PA of \(-20\pm3^\circ\). The energy dependence of polarization is shown in Table~\ref{energypd} and Figure~\ref{energyqi}. 

The PD shows an increasing trend with energy, reaching $14.2\pm2.6\%$ in the 6--7~keV band, which is a rare high value among NS-LMXBs (see, e.g., the review by \citep{Ursini2024TheIV}). A Pearson correlation analysis indicates a positive correlation between PD and energy at a confidence level of $77.3\%$ (correlation coefficient $r = 0.581$, $p = 0.227$), but it is not statistically significant. This is mainly because the PD values in the 2--3, 3--4 and 4--5~keV bands do not show significant differences between themselves (differences $<1\sigma$), while they differ considerably from the PD values above 5~keV. Meanwhile, the PD values in the 5--6, 6--7, and 7--8~keV bands also do not show significant differences between themselves (differences $<1.3\sigma$). Therefore, we separately derived the PD values for the 2--5~keV and 5--8~keV bands, obtaining $5.6\pm0.8\%$ and $11.6\pm1.7\%$, respectively, which differ at a confidence level of approximately $99.9\%$ ($>3\sigma$). Furthermore, the PA does not show an energy-dependent trend.

We divided the observation into ten equal time intervals, each of 1 day, to study the time-resolved polarization properties. The results are presented in Figure \ref{time_pdpa}. Both PD and PA exhibit some variability. Over four days (MJD 60953.5--60957.5), PD increased from $4.1\pm3\%$ (consistent with zero) to $13.3\pm3\%$, with a significance of $2.17\sigma$ ($p\approx0.03$), providing moderate/marginal evidence for an increase in PD. Over three days (MJD 60954.5--60957.5), PA rotated from $-44\pm10^\circ$ to $-5\pm7^\circ$, with a significance of $3.19\sigma$ ($p\approx0.0014$), indicating a significant rotation of $\sim40^\circ$ in PA. However, the corresponding X-ray light curve did not show significant changes during this period. In addition, due to the large uncertainties in the results for the divided energy bands (2--5~keV and 5--8~keV), we do not plot them in the figure, but the polarization trends in these two bands are consistent with the full band.

We also investigated the polarization properties at different orbital phases. First, we divided the observation into three unequal intervals according to the features of the orbital light curve: pre-eclipse (phase 0.00--0.45, corresponding to the broad dip before the companion eclipse), eclipse (phase 0.45--0.58, corresponding to the narrow dip caused by the companion eclipse), and post-eclipse (phase 0.58--1.00, corresponding to the rising phase after the eclipse). In the 2--8 keV band, the PD values in the three phase intervals are broadly consistent: $7.1\pm1.4\%$ (pre-eclipse), $10\pm3.2\%$ (eclipse), and $8.2\pm1.3\%$ (post-eclipse). Moreover, the PA values in the three intervals are also consistent ($\sim-20^\circ$).

Given the significant difference in polarization between the 2--5 keV and 5--8 keV bands, we further performed an energy-resolved analysis for these three phase intervals. The results for the 2--5~keV and 5--8~keV bands are summarized in Table \ref{tab:orbital}. The PD in the two bands shows different trends. In the 2--5~keV band, the PD during the non-eclipse is $\sim5.9\pm0.8\%$, significantly non-zero at $>99.9\%$ confidence ($7.4\sigma$). However, during the eclipse the PD drops to $2.4\pm2.0\%$, consistent with zero ($1.2\sigma$, not significant). In contrast, in the 5--8~keV band, the PD during the non-eclipse is $\sim10\pm1.8\%$, while during the eclipse the PD reaches as high as $21.3\pm6.1\%$, showing an increasing trend compared to the non-eclipse ($1.64\sigma$, marginally significant).

\begin{table*}
\centering
\begin{tabular}{lcccc}
\toprule
\hline
& \multicolumn{2}{c}{\textbf{2--5 keV}} & \multicolumn{2}{c}{\textbf{5--8 keV}} \\
\hline
\textbf{Interval}& \multicolumn{1}{c}{\textbf{PD (\%)}} & \multicolumn{1}{c}{\textbf{PA (deg)}} & \multicolumn{1}{c}{\textbf{PD (\%)}} & \multicolumn{1}{c}{\textbf{PA (deg)}} \\
\midrule
Pre-eclipse  & $5.9\pm1.2$ & $-15\pm6$  & $9.6\pm2.7$ & $-20\pm3$  \\
Eclipse      & $2.4\pm2.0$ & --         & $21.3\pm6.1$ & $-18\pm8$         \\
Post-eclipse & $6.0\pm1.1$ & $-20\pm3$  & $11.9\pm2.5$ & $-22\pm5$  \\
\bottomrule
\end{tabular}
\caption{PD and PA for the three parts of the orbit (pre-eclipse, eclipse, and post-eclipse) obtained by PCUBE algorithm.}
\label{tab:orbital}
\end{table*}

\begin{figure*}
\centering
\includegraphics[width=0.92\textwidth]{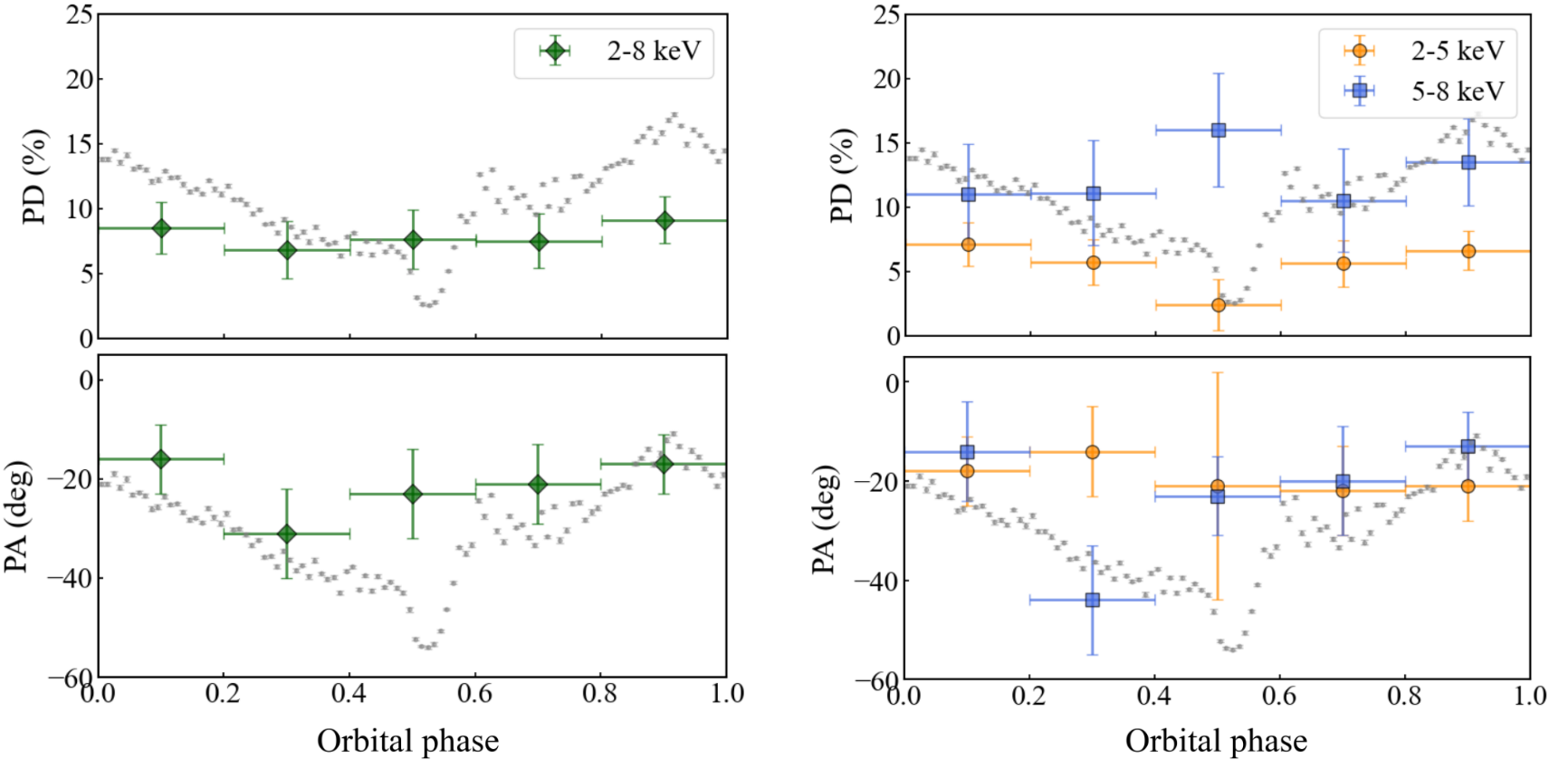}
\caption{Orbital phase dependence of polarization for 4U 1822-37 obtained with the PCUBE algorithm. Left panels: PD (top panel) and PA (bottom panel) in the 2--8~keV band. Right panels: PD (top panel) and PA (bottom panel) in the 2--5~keV and 5--8~keV bands. The gray points show the profile of IXPE orbital light curve.}
\label{ob_pdpa}
\end{figure*}

We also binned the data into equal orbital phase intervals of 0.2. The phase dependence of PD and PA in the 2--8 keV, 2--5 keV, and 5--8 keV bands is shown in Figure \ref{ob_pdpa}. The PD variations are consistent with the trends obtained by using the non-uniform phase division. The PA reveals more detailed orbital variability. In the 2--5 keV band, the PA remains unchanged with orbital phase. However, in the 5--8 keV band, the PA rotates from $-14\pm10^\circ$ at the beginning of the broad dip (phase interval 0.0--0.2) to $-44\pm10^\circ$ just before the companion eclipse (phase interval 0.2--0.4), supporting a rotation of $30^\circ$ at a confidence level of \(\sim\) $96.6\%$. Subsequently, the PA returns to $\sim-20^\circ$ at the companion eclipse and remains roughly at this level in the following phase intervals.

\subsection{Spectral analysis} \label{Spectral analysis}

Fits to data from RXTE/ASCA, EXOSAT, and BeppoSAX show that the continuum emission of 4U 1822-37 above 2~keV can be described by a combination of a blackbody with a temperature of 1.3--2~keV and a flat power-law (or Comptonization) component \citep{1989MNRAS.239..715H, Heinz2000TheFD, Parmar2000BroadbandBO}. The BeppoSAX spectral fitting in the 0.3--40~keV band indicates that the blackbody contributes 40\%--50\% of the total flux in the 1--10~keV range, with its peak contribution in the 2--5~keV band \citep{Parmar2000BroadbandBO}. Moreover, \cite{1989MNRAS.239..715H} found that the blackbody flux implies an emission area consistent with 1/400 of the neutron star surface area based on EXOSAT data, and therefore suggested that this radiation originates from the neutron star surface and is scattered into our line of sight by a very low optical depth corona. In contrast to the blackbody, the flat power-law or Comptonized component has a harder spectrum, peaking at higher energies. However, the spectral shapes of the two components are very similar in the 2--10~keV range, so that we cannot separate them using different spectral models with IXPE and Swift-XRT.

Below 2~keV, observations with the Einstein Solid State Spectrometer and the Suzaku satellite also indicate the presence of a soft excess, which can be modeled as a hot bremsstrahlung with $kT\sim0.25$ keV, a blackbody with $kT\sim0.16$ keV, or an emission feature with an equivalent width of 350 eV at 0.8 keV \citep{White1981AccretionDC, Sasano2013SuzakuVO}. This may represent Fe L complex emission or an additional thermal component.

We fit the Swift-XRT spectra of the five observations using \texttt{tbabs \(\times\) (bremss + powerlaw)} , where \texttt{bremss} is used to model the soft excess below 2~keV. Absorption was modeled with \texttt{tbabs}, which is based on the abundances reported in \cite{2000ApJ...542..914W}. This model is also modified by an absorption edge to describe the decline above $\sim$ 7~keV caused by ionized absorption \citep{1997xisc.conf..411W}. The fitting results are presented in Table~\ref{swift_para}. The derived bremsstrahlung temperature $kT$ and the small photon index $\Gamma$ are consistent with previous studies \citep{Sasano2013SuzakuVO}. In addition, the fifth observation can only be well fitted with an absorbed power-law. Due to possible parameter degeneracy, we do not directly compare its parameters with those of the other observations.

\begin{figure}  %
\centering
\includegraphics[width=0.475\textwidth]{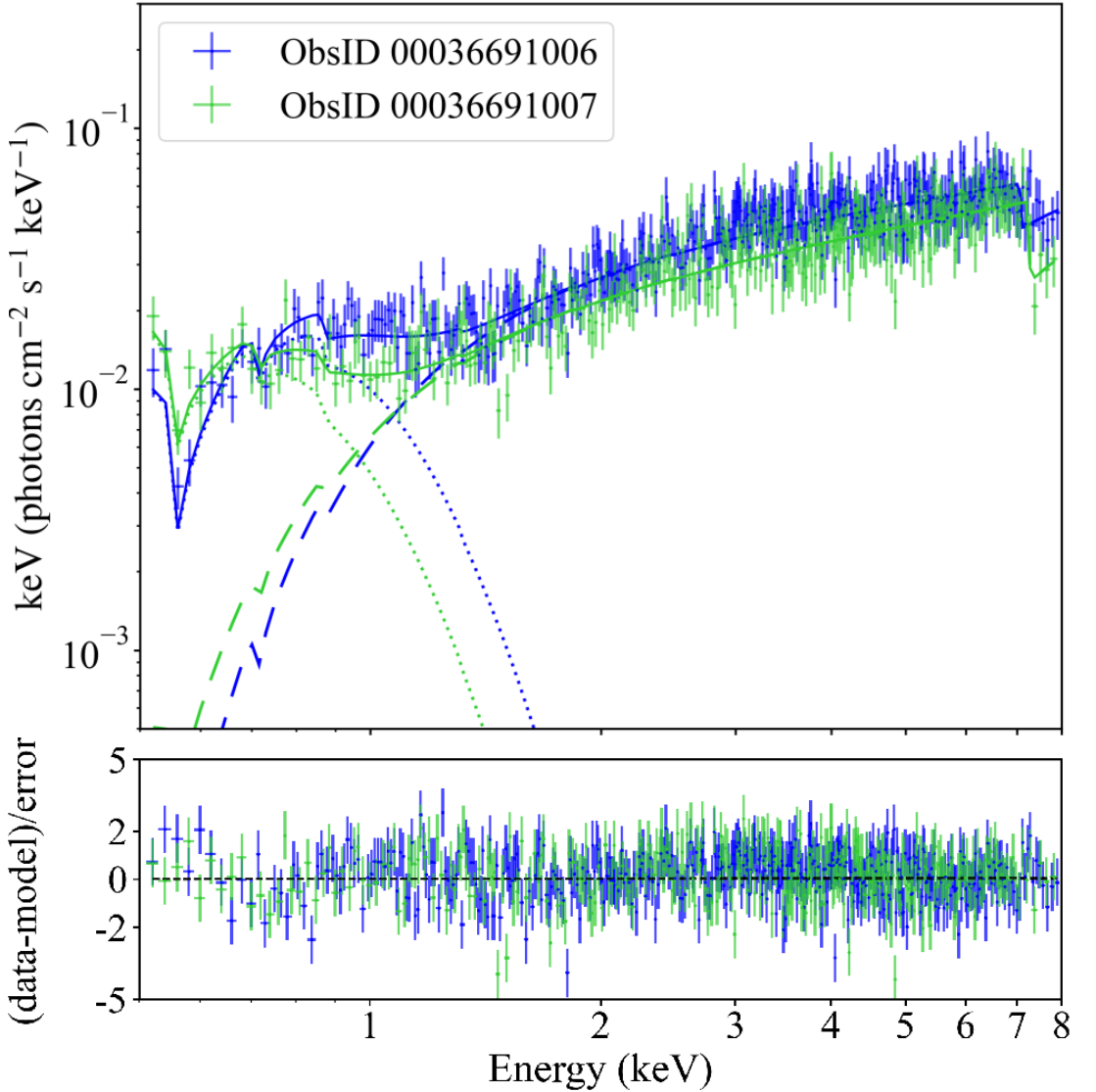}\\
\caption{Spectral energy distribution of the third (ObsID 00036691006) and the fourth (ObsID 00036691007) Swift-XRT observations. The different spectral model components are reported as dotted lines for \texttt{bremss}, dashed lines for \texttt{powerlaw}, and solid lines for the total model. The bottom panel shows the residuals between the data and the best fit.}
\label{swift_spec}
\end{figure}

We note that the time interval corresponding to the PA rotation (PA $\sim-44^\circ$) overlaps with the third Swift-XRT observation, while the interval with PA $\sim-5^\circ$ overlaps with the fourth Swift-XRT observation. The column density of the third observation is significantly higher than that of the fourth observation (as well as the other observations except the fifth one). Besides, the bremsstrahlung temperature $kT$, the power-law index $\Gamma$, and the absorption edge parameters show no significant changes between the third and fourth observations. The difference in spectral shape between the two is shown in Figure~\ref{swift_spec}.

\begin{figure}  %
\centering
\includegraphics[width=0.48\textwidth]{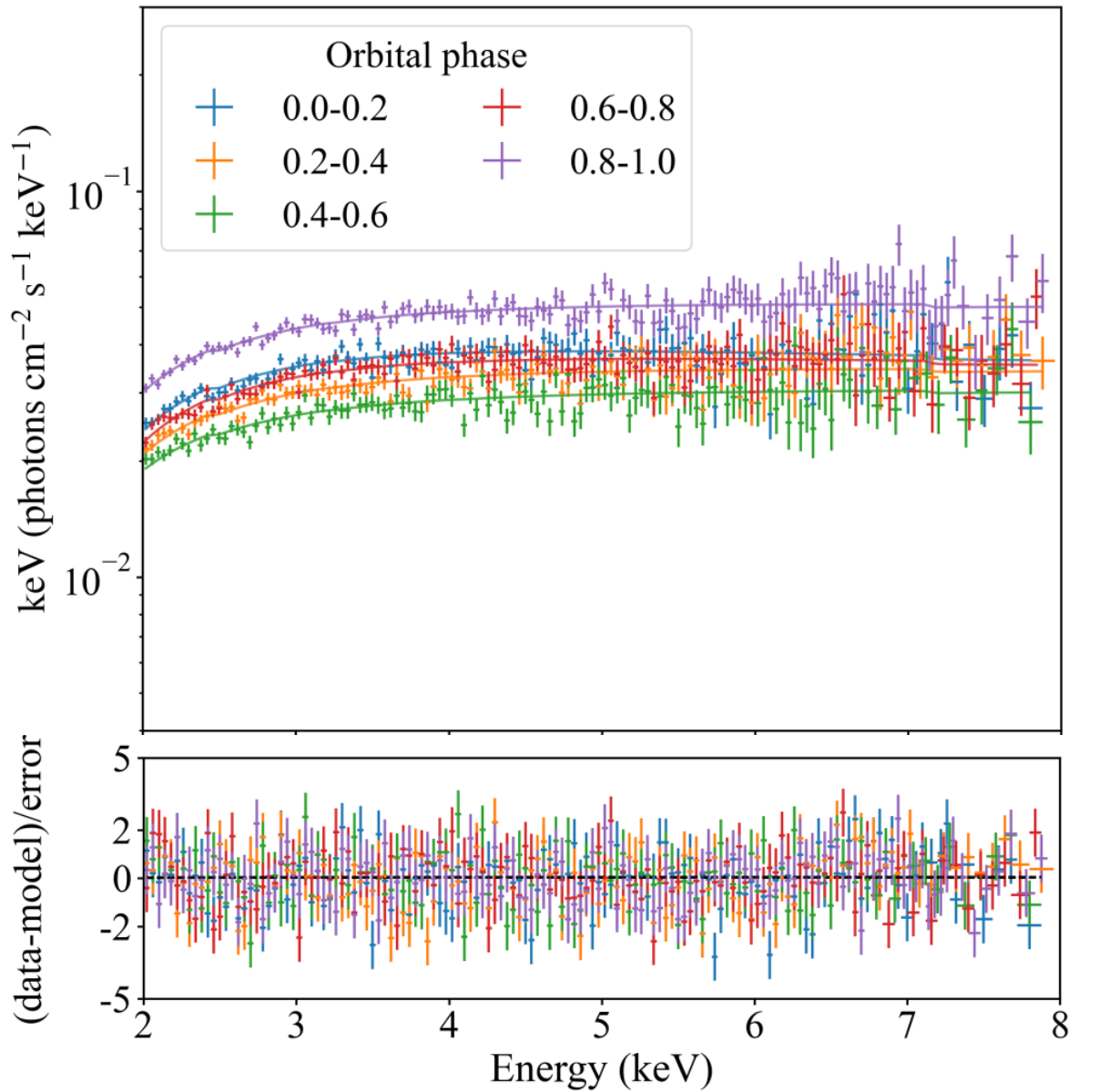}\\
\caption{Spectral energy distribution of five equal orbital phase intervals of the IXPE observation. Here, only DU 1 is reported as an example.}
\label{ixpe_spec}
\end{figure}

We fit the IXPE spectra with \texttt{tbabs \(\times\) powerlaw \(\times\) constant}, where the \texttt{constant} component accounts for possible cross-calibration differences among the IXPE DUs. Given the limited spectral capability of IXPE and its small effective area above 6~keV, we did not attempt to include the absorption edge that is clearly present in the Swift-XRT spectra. The fitting results are presented in Table~\ref{ixpe_para} and Figure \ref{ixpe_spec}. This model provides a good fit to the IXPE spectra, with $\chi^2/{\rm d.o.f}=359/295$. Replacing \texttt{powerlaw} with \texttt{diskbb} also yields an acceptable fit ($\chi^2/{\rm d.o.f}=378/295$). However, the inner disk temperature parameter in \texttt{diskbb} reaches $\sim5$ keV, an unphysical value implying that this component merely provides a mathematical approximation of the power-law shape. Moreover, replacing the \texttt{powerlaw} with a \texttt{bbody} results in a completely unacceptable fit ($\chi^2/\mathrm{d.o.f}=1489/295$).

As shown in Table~\ref{ixpe_para}, the photon index varies between 1.01 and 1.13 in the phase range 0.2--0.8. The column density during the eclipse (phase 0.4--0.6) is significantly lower than that during the post-eclipse (phase 0.6--0.8), consistent with the companion star obscuring the central absorber.

\begin{table*}
\centering
\renewcommand{\arraystretch}{1.3}  
\begin{tabular}{l l c c c c c}
\toprule
\hline
\textbf{Component} & \textbf{Parameter} & $\mathbf{00036691004}$ & $\mathbf{00036691005}$ & $\mathbf{00036691006}$ & $\mathbf{00036691007}$ & $\mathbf{00036691008}$ \\
\midrule
\multirow{1}{*}{tbabs} & $N_\mathrm{H}$ ($10^{22}$ cm$^{-2}$) & $0.37^{+0.09}_{-0.08}$ & $0.66\pm0.10$ & $0.80\pm0.11$ & $0.56\pm0.10$ & $2.09^{+0.46}_{-0.45}$ \\
\midrule
\multirow{2}{*}{bremss} & kT (keV) & $0.16^{+0.02}_{-0.01}$ & $0.11\pm0.01$ & $0.17^{+0.02}_{-0.01}$ & $0.15^{+0.02}_{-0.01}$ & $-$ \\
                         & Norm & $1.36^{+1.60}_{-0.64}$ & $18.42^{+25.86}_{-10.67}$ & $7.13^{+6.98}_{-3.67}$ & $4.16^{+5.05}_{-2.33}$ & $-$ \\
\midrule
\multirow{2}{*}{powerlaw} & $\Gamma$ & $0.42^{+0.05}_{-0.04}$ & $0.50\pm0.05$ & $0.52\pm0.05$ & $0.44\pm0.05$ & $0.79\pm0.08$ \\
                          & Norm & $0.021\pm0.001$ & $0.017\pm0.001$ & $0.024\pm0.002$ & $0.017\pm0.001$ & $0.034\pm0.005$ \\
\midrule
\multirow{2}{*}{edge} & $E$ (keV) & $6.99^{+0.14}_{-0.16}$ & – & $7.13^{+0.36}_{-0.33}$ & $7.25^{+0.13}_{-0.12}$ & $-$ \\
                         & $\tau$ & $0.54^{+0.12}_{-0.10}$ & – & $0.40^{+0.19}_{-0.12}$ & $0.72^{+0.22}_{-0.17}$ & $-$ \\
\midrule
&$\chi^2$/d.o.f & 427/432 & 266/257 & 385/379 & 349/291 & 299/238 \\
\bottomrule
\end{tabular}
\caption{Spectral fitting results using \textsc{xspec} with the \texttt{edge \(\times\) tbabs \(\times\) (bremss + powerlaw)} model for all five Swift-XRT observations (ObsID 00036691004 -- 00036691008). Uncertainties are 1$\sigma$ confidence intervals from the \texttt{error} command. For ObsID 00036691005, the edge parameters were not required (the fit with a simple model is shown, the edge component was omitted).}
\label{swift_para}
\end{table*}

\subsection{Spectropolarimetric analysis} \label{Spectropolarimetric analysis}

Since the model-independent polarimetric analysis using PCUBE in Section~3.1 revealed an energy-dependent trend in which the PD in the high-energy band is significantly higher than that in the low-energy band, we convolved the powerlaw model in the IXPE fitting model with \texttt{polconst} and \texttt{pollin}, respectively, to further test this energy dependence.

\texttt{polconst} assumes a constant polarization, yielding \(\chi^2/\mathrm{d.o.f}=958/889\), with \(\mathrm{PD}=5.8\pm0.5\%\) and \(\mathrm{PA}=-20\pm3^\circ\), which are consistent with the model‑independent PCUBE results within \(3\sigma\). \texttt{pollin} describes an energy‑dependent linear polarization, \(\mathrm{PD} = A_1 + A_{\mathrm{slope}}(E - E_0)\), with \(E_0 = 1\)~keV, and we additionally assumed that the PA is energy independent. For the 2--8~keV band, \texttt{pollin} gives \(\chi^2/\mathrm{d.o.f}=945/888\), which is a significant improvement over \texttt{polconst} with \(\Delta\mathrm{AIC}=11\) (where AIC is the Akaike Information Criterion). The best fit gives \(A_1\) consistent with zero and \(A_{\mathrm{slope}} = 0.017 \pm 0.005\), indicating that PD increases linearly with energy. The PD values derived from \texttt{pollin} agree with the PCUBE results within \(3\sigma\) (e.g., \texttt{pollin} gives a PD of \(8.4\pm2.5\%\) at 5.5~keV, while PCUBE gives \(12.7\pm1.9\%\) in the 5--6~keV band).

However, the \texttt{pollin} fit in the 2--5~keV band yields \(A_1 = 0.045 \pm 0.020\) and a slope \(A_{\mathrm{slope}} = 0.0004 \pm 0.0086\) (consistent with zero), implying a constant PD of \(4.5\pm2.0\%\). For the 5--8~keV band, the fit gives \(A_1 = 0.156 \pm 0.095\) and \(A_{\mathrm{slope}} = -0.010 \pm 0.019\) (also consistent with zero), implying a constant PD of \(15.5\pm9.5\%\). Moreover, the reduced \(\chi^2\) values from the \texttt{pollin} fits in the divided energy bands are consistent with those from \texttt{polconst} (e.g., in the 2--5~keV band, \texttt{pollin} gives \(\chi^2/\mathrm{d.o.f}=455/438\), while \texttt{polconst} gives \(455/439\)). Although the uncertainties on the PD in the sub‑bands are large, the energy‑independent behavior of the slope indicates that the full‑band linear polarization trend derived from \texttt{pollin} likely arises from the difference in the mean PD values between the low‑ and high‑energy bands. This result is identical to that obtained with PCUBE.

In addition, we performed similar energy-dependent analyzes with \texttt{pollin} for the different time intervals and orbital phase intervals defined in Section 3.1 and found no significant changes in the energy-dependent properties compared to the full dataset.

\section{DISCUSSION} \label{sec:discussion}

\subsection{Polarimetric properties of ADC}

We report the first X-ray polarimetric measurement of 4U 1822-37. Its PD remains constant within the 2--5~keV and 5--8~keV bands, with average values of \(5.6\pm0.8\%\) and \(11.6\pm1.7\%\), respectively, and the difference is significant (\(>3\sigma\)). In addition, PA in the two energy bands are generally consistent. However, the two bands exhibit distinctly different polarization behaviors as a function of orbital phase, clearly indicating that the polarization processes or source regions in the 2--5 keV and 5--8 keV bands may be different.

Although the broadband spectral fit with IXPE in the 0.4--0.6 phase interval does not show a significant variation in the photon index \(\Gamma\), the finer hardness ratio drops sharply from a high value at the shoulder (phase 0.4--0.5) to a low value at the eclipse center (phase 0.5--0.6), by about 28\%. This is consistent with previous RXTE and ASCA observations, which showed that the fractional depth of the companion's eclipse increases with energy from sub-keV to \(\sim7\)~keV \citep{Heinz2000TheFD}. This implies that the 5--8~keV emission originates from regions that are generally closer to the disk plane than the 2--5~keV emission, and thus is more severely occulted during the eclipse.

The 5--8~keV band exhibits a very high PD ($\sim12\%$), reaching as high as $\sim20\%$ during the eclipse, which is much higher than the typical 1--5\% seen in most LMXBs \citep{Ursini2024TheIV}. The high PD in LMXBs may be related to scattering in disk winds or Comptonization in outflows, with possible contributions from reflection components (see, e.g., \cite{2025arXiv251213182M, DiMarco2023FirstDO}).

\begin{table*}
\centering
\renewcommand{\arraystretch}{1.3} 
\begin{tabular}{l l c c c c c c}
\toprule
\hline
\textbf{Component} & \textbf{Parameter} & \textbf{Phase-averaged} & \textbf{0.0-0.2} & \textbf{0.2-0.4} & \textbf{0.4-0.6} & \textbf{0.6-0.8} & \textbf{0.8-1.0} \\
\midrule
\multirow{1}{*}{tbabs} & $N_\mathrm{H}$ ($10^{22}$ cm$^{-2}$) & $2.07 \pm 0.06$ & $2.39 \pm 0.13$ & $1.99 \pm 0.14$ & $1.74 \pm 0.15$ & $2.31 \pm 0.14$ & $2.08 \pm 0.12$ \\
\midrule
\multirow{2}{*}{powerlaw} & $\Gamma$ & $1.06 \pm 0.01$ & $1.17 \pm 0.03$ & $1.04 \pm 0.03$ & $1.01 \pm 0.04$ & $1.13 \pm 0.03$ & $1.05 \pm 0.03$ \\
                          & Norm & $0.044 \pm 0.001$ & $0.054 \pm 0.003$ & $0.038 \pm 0.002$ & $0.032 \pm 0.002$ & $0.048 \pm 0.003$ & $0.057 \pm 0.003$ \\
\midrule
\multirow{2}{*}{constant} & DU 1 & [1.000] & [1.000] & [1.000] & [1.000] & [1.000] & [1.000] \\
                         & DU 3 & $1.004 \pm 0.003$ & [1.004] & [1.004] & [1.004] & [1.004] & [1.004] \\
\midrule
&$\chi^2$/d.o.f & 359/295 & 313/271 & 286/269 & 302/261 & 296/269 & 276/281 \\
\bottomrule
\end{tabular}
\caption{Spectral fitting results using \textsc{xspec} with the \texttt{tbabs \(\times\) powerlaw \(\times\) constant} model for IXPE spectra. The phase-averaged spectrum and the phase-resolved spectra (phase intervals 0.0–0.2, 0.2–0.4, 0.4–0.6, 0.6–0.8, 0.8–1.0) are shown. For the phase-resolved spectra, the \texttt{constant} factors for the different DUs are fixed to their phase-averaged values. Uncertainties are 1$\sigma$ confidence intervals from the \texttt{error} command.}
\label{ixpe_para}
\end{table*}

\citet{Anitra2026TheFI} used the same IXPE observation combined with Swift-XRT, XMM-Newton and NuSTAR, and detected a broad reflection component whose polarization parameters could not be reliably constrained, so its intrinsic PD was ultimately fixed to 15\% and 20\% in the fitting. However, this component accounts for only \(\sim11\%\) of the flux in the 6--7~keV band, and has an average fraction of only \(\sim2-6\%\) in other bands. Therefore, such a low-fraction reflection component alone cannot explain the overall high PD in the 5--8~keV band, nor the significant PD difference between the 2--5~keV and 5--8~keV bands. Even when considering the polarization from an optically thick Comptonizing slab, the high PD of 10--20\% cannot be easily explained, since for an inclination of \(i\sim80^\circ\), the maximum PD from an optically thick Comptonizing slab is only \(\sim8\%\) \citep{1985A&A...143..374S}. Thus, the 5--8~keV emission most likely includes a scattering contribution from an extended optically thin medium (possibly related to the outer ADC or a disk wind), with the seed photons likely dominated by the Comptonizing region rather than the soft thermal radiation from the disk, given that the \texttt{bbody} or \texttt{diskbb} models provide poor fits. At high inclinations, such scattering along with possible Comptonized direct radiation from optically thick regions can produce a net PD greater than 15\% \citep{Nitindala2024XrayPF}.

The 5--8~keV PD remains unchanged or shows a marginal rising trend during the eclipse. If the main contributing region of the hard polarization were fully occulted by the companion during the eclipse, the PD would be expected to drop significantly. Therefore, the observed PD behavior implies that the companion eclipse may only cover regions that contribute little to the hard polarization, or that the polarizing region itself has a certain extent, for example along the disk plane, given that the hardness ratio decreases during the eclipse, suggesting that the hard component is vertically confined. In summary, this PD behavior helps to constrain the geometry of the scattering medium.

In other high-inclination LMXBs, a rising trend of PD during eclipse has also been observed, often attributed to the partial occultation of the direct radiation by the companion, which reduces its flux fraction, while the fraction of the highly polarized scattered radiation (mainly from the disk wind and/or corona) increases, leading to a rise in net PD. For example, the high-inclination source AX~J1745.6-2901 shows an increase in PD from \(\sim9\%\) outside the eclipse to \(\sim34\%\) during the eclipse \citep{2025arXiv251213182M}. In addition, the high-inclination ADC source 2S~0921-630 also exhibits a PD increase during eclipse, from \(\sim6\%\) to 15\%, with its high PD attributed to disk wind scattering \citep{Tomaru2025TheDO}. However, it is worth noting that these sources show a significant PD rise, in contrast to the marginal rise of the hard PD in 4U~1822--37, suggesting that we should be cautious in interpreting the PD variations in 4U~1822--37 as due to changes in the scattering-to-direct ratio.

The polarization behavior in the 2--5~keV band suggests that its radiation geometry differs from that of the 5--8~keV band. The 2--8~keV spectrum does not support the \texttt{diskbb} model, so we do not consider a significant contribution from the disk. Previous broadband observations of 4U~1822--37 detected a blackbody component in the 1--10~keV band, with an emission area consistent with 1/400 of the neutron star surface area, and it contributes significantly in the 2--5~keV band \citep{Heinz2000TheFD, Parmar2000BroadbandBO}. However, the source shows almost no pulsations in the soft band. Non-pulsating isotropic thermal radiation or Comptonization in structures near the neutron star (e.g., spreading layers or boundary layers; \cite{Inogamov1999SpreadOM, Suleimanov2006SpectraOT, Popham2000AccretionDB}) also cannot easily explain the 5--6\% PD outside the eclipse. These difficulties suggest that scattering may be a viable direction to consider \citep{Nitindala2024XrayPF}.

The behavior of the 2--5~keV PD during the eclipse further supports this possibility. The observed 2--5~keV radiation may be consistently dominated by scattering, with almost no direct component; otherwise it would be difficult to explain why the PD drops to a level consistent with zero during the eclipse, rather than increasing as in the case of high-inclination sources like AX~J1745.6-2901, where the direct-to-scattered ratio decreases due to the companion occultation. A natural interpretation is that the eclipse mainly occults the scattering region that contributes significantly to the soft polarization, and the remaining scattered signal is insufficient to maintain a detectable PD. This is consistent with the picture in high-inclination ADC sources where the outer disk significantly blocks the central direct emission, suggesting that the soft direct radiation likely mainly originates from a central compact region, such as the neutron star surface.

Overall, the significantly different PD in the 2--5~keV and 5--8~keV bands support a difference in the spatial distribution of the polarization sources between the two bands. One possibility is that the soft polarization mainly arises from scattering of radiation from the neutron star surface or a nearby compact region, while the hard polarization mainly arises from scattering in a more extended region along the disk plane. The former is scattering-dominated due to occultation by the outer disk, while the latter may contain a certain fraction of direct radiation (e.g., Comptonization near the disk plane). This picture is compatible with the scattering-dominated scenario proposed by \cite{Anitra2026TheFI}, while our energy-resolved analysis further reveals that the scattering contribution exhibits different spatial distributions and geometric properties between the soft and hard bands.

As a possible physical picture, the magnetic field of 4U~1822--37, which is orders of magnitude stronger than that of typical LMXBs, may be relevant to this. In the strong magnetic field case, the accretion flow follows magnetic field lines toward the neutron star magnetic poles at the magnetospheric radius, and the Comptonizing region near the disk cannot fully cover the neutron star. Radiation from different regions may experience different scattering and occultation geometries, giving rise to the distinct polarization behavior between the soft and hard bands. In contrast, another weakly magnetized ADC source observed by IXPE, 2S~0921-630, does not show significant energy-dependent PD variations \citep{Tomaru2025TheDO}. Suzaku observations indicate a massive (\(R\sim10^9\)~cm) optically thick Compton cloud between the neutron star and the inner disk \citep{2023PASJ...75...30Y}, in which most of the original X-ray radiation is reprocessed \citep{2003ApJ...583..861K}. This may be related to the absence of significant energy dependence of PD in this source \citep{Tomaru2025TheDO}.

\subsection{PA variability caused by asymmetric structures}

PA of 4U~1822--37 exhibits two types of variation. The orbital phase-resolved analysis reveals that the 5--8~keV PA rotates periodically by about \(30^\circ\) over phases 0.2--0.4 (i.e., before the companion eclipse), whereas the 2--5~keV PA remains nearly constant and no significant variations in spectral parameters are observed. In addition, on a three-day timescale, PA rotates by approximately \(40^\circ\), accompanied by an increase in neutral absorption but no significant variability in X-ray luminosity.

The orbitally modulated PA variation can likely be related to the orbital light curve. Before the eclipse, the light curve shows a broad dip centered near phase 0.4, usually interpreted as obscuration of the central emission by a bulge formed by the accretion stream impacting the outer disk edge (e.g., \cite{White1981AccretionDC,1989MNRAS.239..715H}). \cite{Cottam2001HighResolutionXS} used Chandra/HETGS to detect photoionized recombination emission and, based on its phase dependence, associated its emission region with the bulge. From the emission measure of the recombination lines, they inferred that this region may have a high electron density (e.g., for material within a Thompson depth of \(\tau=1\), \(n_e \sim 10^{11}\,\mathrm{cm^{-3}}\)). Recently, \cite{Sameshima2026XrayDT} performed Doppler tomography and constructed a velocity map of the Fe K\(\alpha\) line, ultimately locating the line near the accretion stream–disk overflow, i.e., in the vicinity of the aforementioned bulge. This suggests that a substantial amount of cold material may be present at the bulge.

Although the low flux fraction of the 6.4~keV Fe K\(\alpha\) line in the continuum is insufficient to significantly affect the 5--8~keV polarization, the dense material in the bulge region itself may alter the radiation geometry along the line of sight and thus contribute to the polarization variations. For example, occultation by the bulge may change the projected geometry of scattering media such as the corona and/or disk wind along the line of sight, leading to a net rotation of PA. Similar interpretations have also been applied to other high-inclination sources. For example, in the ADC source 2S~0921-630, an energy-dependent PA swing of \(\sim35^\circ\) (from 2--3~keV to 5--8~keV) was detected, and the non-axisymmetry of the outer disk caused by the bulge is considered as a possible explanation \citep{Tomaru2025TheDO}. For the dipping source GX~13+1, where the PA swings by \(70^\circ\) with no significant spectral changes \citep{Bobrikova2024DiscoveryOA}. After joint fitting with NICER, an increase in the Comptonized component was found, which was associated with a possible bulge or clump occultation \citep{DiMarco2025XRayDA}. The behavior of 4U 1822-37 is analogous, though more multi‑wavelength data are needed to detect potential high‑energy spectral changes.

Since the PA variation is not accompanied by a significant change in PD, the fraction of the direct radiation is unlikely to change substantially, further supporting the idea that this effect is dominated by changes in the scattering geometry. In addition, the fact that only the 5--8~keV PA changes while the 2--5~keV PA remains constant may reflect a difference in the scattering geometry between the two bands. For example, the scattering region in the 2--5~keV band may be diffuse in all directions, so that the occultation by the bulge does not produce a significant geometric asymmetry with azimuth. In contrast, the 5--8~keV scattering region may be more compact in the vertical direction but extended in other directions (e.g., along the disk plane), and the occultation by the bulge could significantly alter the geometric asymmetry of the scattering medium, leading to a net PA rotation.

Additionally, the \(\sim40^\circ\) rotation of PA on a three-day timescale, accompanied by an increase in neutral absorption with no change in luminosity, may indicate a slow evolution of the overall structure (thickness or geometry) of the bulge or other absorbers along the line of sight. Such an evolution would likely alter the weight of the emission that undergoes different absorption/scattering processes, causing the PA to rotate while the total flux shows little variability. Since this timescale is much longer than the orbital period, the PA variation involves the evolution of the absorber itself rather than the orbital geometry, and it is difficult to further distinguish the specific mechanism with the current observations alone. Therefore, these scenarios require more detailed modeling in the future.


\section{Summary}
\label{sec:summary}

Based on the IXPE and Swift-XRT observations of 4U~1822--37, we have obtained the first X-ray polarimetric results for this source. The main findings are summarized as follows.

\begin{enumerate}
\item In the 2--8~keV band, 4U~1822--37 exhibits an average PD of \(7.8\pm0.9\%\) and an average PA of \(-20\pm3^\circ\). The PD is significantly higher than the typical values of 1\%--5\% found in classical LMXBs, likely suggesting a scattering contribution from an extended optically thin medium (possibly related to the outer ADC or a disk wind).

\item The PD shows a significant energy dependence. In the 2--5~keV band, the PD is \(5.6\pm0.8\%\), while in the 5--8~keV band it is \(11.6\pm1.7\%\), with a difference exceeding the \(3\sigma\) confidence level. The PA, however, remains consistent between the two bands.

\item During the companion eclipse, the PD in the 2--5~keV band drops from a non-zero value (\(\sim5.9\%\)) to near zero (\(2.4\pm2.0\%\)). In contrast, the PD in the 5--8~keV band shows a marginal rising trend during the eclipse, reaching \(21.3\pm6.1\%\). These findings support different radiation geometries or dominant mechanisms between the two energy bands.

\item Before the companion eclipse, the PA in the 2--5~keV band remains constant. In contrast, the PA in the 5--8~keV band undergoes a periodic rotation of \(\sim30^\circ\) before the eclipse. This rotation may be related to changes in the scattering geometry caused by the bulge at the accretion stream--disk impact.

\item On a timescale of about three days, the PA rotates by \(\sim40^\circ\), accompanied by an increase in the neutral absorption column density but no significant change in X-ray luminosity. This phenomenon may reflect variations in the absorbing material distribution along the line of sight, or a slow evolution of the bulge structure at the outer edge of the accretion disk.

Overall, the polarimetric results of 4U~1822--37 support a picture in which the polarization sources in the soft and hard bands have different spatial distributions. The 2--5~keV radiation is likely scattering-dominated; during the eclipse, the companion partially occults the scattering region, and the remaining region is insufficient to maintain a detectable polarization signal. Its direct emission likely originates from a central compact region (e.g., the neutron star surface) that is fully blocked by the outer disk. The 5--8~keV band may contain direct emission from a more extended region along the disk plane, as well as scattering contributions. The magnetic field of this source, which is orders of magnitude stronger than that of typical LMXBs, may be related to this difference in spatial distribution.

\end{enumerate}

\section*{Acknowledgments}

This work is supported by National Natural Science Foundation of China (grant No. 12422306), and Natural Science Foundation of Guangxi (grant Nos. 2025GXNSFDA02850001), and Bagui Scholars Program (XF). 
This work is also supported by the Guangxi Science and Technology Innovation Platform Program (Leitai Action Plan, Grant No. Guike LT2600640026), Guangxi Key R\&D Program (Guangxi Funeng Action Plan, Grant No. Guike FN2504240040), and the ``Guangxi Highland of Innovation Talents'' Program.
All authors provided inputs and comments on the manuscript. The authors declare no conflicts of interest.

\section*{Data availability}

The data underlying this article are available on the HEASARC website
(\url{https://heasarc.gsfc.nasa.gov/docs/archive.html})



\bibliographystyle{mnras}
\bibliography{sample701} 


\bsp	
\label{lastpage}
\end{document}